\documentclass[reprint,twocolumn]{revtex4-1}      %aps,pra,amsmath,amssymb,{revtex4}
\usepackage{amsmath}
\usepackage{amssymb}
\usepackage{bm}% bold math
\usepackage{color}
\usepackage{multirow}
\usepackage{makecell}
\usepackage{float}
\usepackage{graphicx}
\usepackage{upgreek}
\usepackage{dcolumn}
\usepackage{tikz}
\def\be{\begin{equation}}
\def\ee{\end{equation}}
\def\bea{\begin{eqnarray}}
\def\eea{\end{eqnarray}}

\newcommand{\appropto}{\mathrel{\vcenter{
  \offinterlineskip\halign{\hfil$##$\cr
    \propto\cr\noalign{\kern2pt}\sim\cr\noalign{\kern-2pt}}}}}

\begin{document}

\title{Graphene-hBN resonant tunneling diodes as high-frequency oscillators}
\author{J. Gaskell,$^1$ L. Eaves,$^{1,2}$ K.S. Novoselov,$^2$ A. Mishchenko,$^2$ A.K. Geim,$^{2,3}$ T.M. Fromhold,$^1$ and M.T. Greenaway$^{1}$  \\$^1$School of Physics and Astronomy, University of Nottingham, Nottingham, NG7 2RD, UK \\ $^2$School of Physics and Astronomy, University of Manchester, Oxford Road, Manchester, M13 9PL, UK \\ $^3$Centre for Mesoscience and Nanotechnology, University
of Manchester, Manchester, M13 9PL, UK}

\date{\today}

\begin{abstract}
We assess the potential of two-terminal graphene-hBN-graphene resonant tunneling diodes as high-frequency oscillators, using self-consistent quantum transport and electrostatic simulations to determine the time-dependent response of the diodes in a resonant circuit. We quantify how the frequency and power of the current oscillations depend on the diode and circuit parameters including the doping of the graphene electrodes, device geometry, alignment of the graphene lattices, and the circuit impedances. Our results indicate that current oscillations with frequencies of up to several hundred GHz should be achievable.
\end{abstract}

\maketitle

Resonant tunneling diodes (RTDs) operating at frequencies and output powers of up to 1.4 THz and 10 $\upmu$W have been recently demonstrated\cite{Suzuki2010,Koyama2013,Feiginov2014}. A new addition to the family of devices that exhibit resonant tunneling and negative differential conductance (NDC) is the graphene-based tunnel transistor \cite{Britnell2012,Nat_Nano,Nat_Comm,Kang2015,Feenstra,Georgiou,Fallahazad2015,Brey2014,Vasko2013,Zhao2013,Ryzhii2013,Ryzhii2013a}. In this device, a thin sheet of hexagonal boron nitride (hBN) acts as a potential barrier separating two monolayer graphene electrodes. The NDC arises from the constraints imposed by energy and momentum conservation of Dirac fermions, which tunnel through the hBN barrier when a bias voltage is applied between the graphene electrodes. Peak-to-valley current ratios (PVRs) of 2:1 have been seen at room temperature, with peak current densities of 0.28 $\upmu$A$/\upmu$m$^{2}$ \ \cite{Nat_Nano,Nat_Comm}. When these devices are placed in a resonant circuit and biased in the NDC region, MHz oscillations occur \cite{Nat_Nano}.

\begin{figure}[t!]
  \centering
\includegraphics[width=1\linewidth]{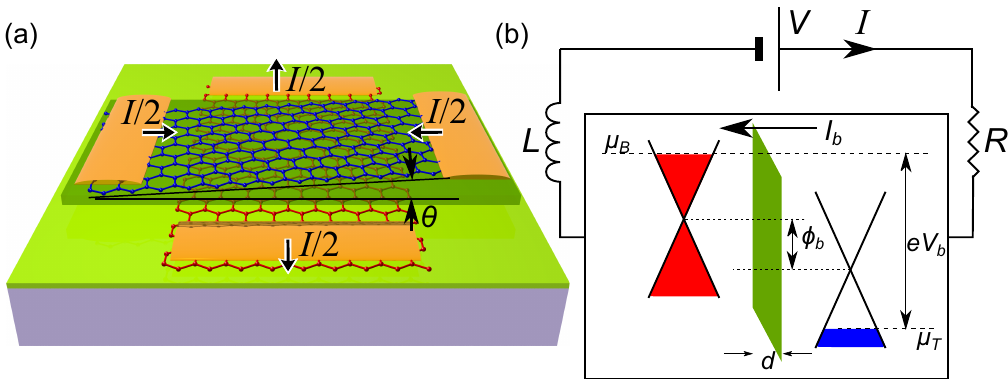}
  \caption{(a) Schematic diagram of the GRTD comprising  bottom (red) and top (blue) graphene lattices, misaligned by an angle $\theta$ and separated by a hBN tunnel barrier (dark green). The current, $I$, passes through the tunnel barrier between the graphene electrode layers to/from Ohmic contacts (orange). The diode is mounted on a hBN layer (light green) and an insulating substrate (purple). (b) Schematic diagram of the resonant circuit incorporating the GRTD (in box) showing the voltage applied, $V$, circuit inductance, $L$, and resistance, $R$. The band diagram is shown (box), with electrostatic parameters defined in the text.} 
\label{fig:1}
\end{figure}

 Here, we use a theoretical analysis to investigate how the device and circuit parameters can be tuned to increase the operating frequency of graphene resonant tunneling diodes (GRTDs). Our model device, shown schematically in Fig. 1(a), comprises two graphene layers separated by a hBN tunnel barrier of thickness, $d$. The bottom (B) and top (T) graphene electrodes are arranged in an overlapping cross formation, resulting in an active tunneling region of area, $A=$ 1 $ \upmu$m$^2$. We consider the general case when the two graphene crystalline lattices are slightly misorientated by a twist angle, $\theta$, see Fig. 1(a). The resonant tunnel current is particularly sensitive to this angle \cite{Nat_Nano}. A bias voltage, $V_b$, applied between the top and bottom graphene layers [Fig. 1(b)] induces a charge density, $\rho_{B,T}$, in each layer and causes a tunnel current, $I_b$, to flow through the hBN barrier. The graphene layers, with in-plane sheet resistance, $R$, carry current, $I$, (black arrows) from two pairs of Ohmic contacts [orange in Fig. 1(a)] to the central active (tunneling) region of the device, i.e. currents, $I/2$, flow to/from each contact.    
The electrostatics of the diode \cite{Britnell2012} are governed by the equation $eV_b=\mu_B-\mu_T+\phi_b$,  where $\phi_b=eF_b d$ is the electrostatic potential energy difference across the barrier, with $F_b$ being the electric field in the barrier, $e$ is the magnitude of the electronic charge, and $\mu_{B,T}$ are the two Fermi levels [see Fig. \ref{fig:1}(b)]. 

A device with NDC can generate self-sustained current oscillations when placed in an RLC circuit \cite{Hines1960}. To investigate the frequency response of the GRTD, we solve the time-dependent current continuity and Poisson equations self-consistently, using the Bardeen transfer Hamiltonian method to calculate the tunnel current,
\be 
I_b=\frac{8\pi e}{\hbar} \sum_{\mathbf{k}_B,\mathbf{k}_T} |M|^2 [f_B (E_B )-f_T (E_T)]\delta(E_B-E_T-\phi_b ), 
\label{eq:currentb}
\ee
as a function of time, $t$, and $V_b$. The summation is over all initial and final states, with wavevectors, $\mathbf{k}_{B,T}$, measured relative to the position of the nearest Dirac point in the bottom layer, $\mathbf{{K}}_{B}^\pm=(\pm 4\pi/3a_0,0)$, where $\pm$ distinguishes the two non-equivalent Dirac points in the Brillouin zone and $a_0=2.46$ \AA \ is the graphene lattice constant. The Fermi function in each electrode is $f_{B,T}(E_{B,T} )=[1+e^{(E_{B,T}-\mu_{B,T})/{kT}} ]^{-1}$ where $E_{B,T}=s_{B,T}\hbar v_F k_{B,T}$ is the electron energy and $s_{B,T}=\pm 1$ labels electrons in the conduction (+) and valence ($-$) bands, at temperature $T=300$ K. Tunneling between equivalent valleys gives the same contribution to the tunnel current, so we consider transitions between $\mathbf{K}^+$ points only. In Eq. (\ref{eq:currentb}) the matrix element, $M$, is 
\be
M=\Xi  \gamma(\theta) g(\varphi_B , \varphi_T) V_S (\mathbf{q}-\mathbf{\Delta K}),%M= \int_{\Omega} d\Omega  \psi^*_{T}  V_S  \psi_{B}
\label{eq:matele}
\ee
where $\Xi$ is a normalisation constant, $\gamma(\theta)$ is the spatial overlap integral of the cell-periodic part of the wavefunction, $g(\varphi_B,\varphi_T)$ describes electron chirality, $V_S$ is the elastic scattering potential, and $\mathbf{q}=\mathbf{k}_B-\mathbf{k}_T$ (see below).  

In recently-studied GRTDs \cite{Nat_Nano}, the crystalline lattices of the two graphene layers are misorientated by only a small twist angle $\theta\approx 1^{\circ }$. Nevertheless, this gives rise to a significant misalignment of the Dirac cones of the two layers, $\Delta\textbf{K}=(R(\theta)-1)\textbf{K}^{+}$, where $R(\theta)$ is the 2D rotation matrix. When $\theta<2^\circ$, $|\mathbf{q}|\approx|\mathbf{\Delta K}|= \Delta K$ and electrons tunnel with conservation of in-plane momentum. However, tunneling electrons can scatter elastically from impurities and defects, broadening the features in the current-voltage curves. Therefore, we use a scattering potential $V_S(q)=V_0/(q^2+{q_c}^2)$, with characteristic lengthscale $1/q_c=15$ nm, which gives the best fit in the region of the resonant peak and NDC \cite{Nat_Comm}. 

The misorientation of the layers also causes the cell-periodic part of the Bloch wavefunctions in the two layers to become misaligned, thus reducing their spatial overlap integral, $\gamma(\theta)$, and, consequently, the tunnel current amplitude. We estimate $\gamma(\theta)$ by calculating the overlap integral of the normalised cell-periodic parts of the Bloch states, $u(\mathbf{r})$, at the Dirac point in the two electrodes over an area, $S_C$, that greatly exceeds the length scale, $\sim 1/q_c$, of the impurity potential, i.e. we take
\be
\gamma(\theta)=\frac{1}{S_C} \int_{S_C} dS_C u^* (R(\theta)\mathbf{r}) u(\mathbf{r}).  %%_{\alpha0}
\ee 

The chiral wavefunctions give rise to the term
\begin{equation}
g(\varphi_B , \varphi_T) = 1 + s_B e^{i\varphi_B}
+ s_T e^{-i\varphi_T} + s_B s_T e^{i(\varphi_B-\varphi_T)}, %\right
\label{eq:gfact}
\end{equation}
where $\varphi= \tan^{-1}(k_y/k_x)$ is the orientation of the wavevector. Finally, in Eq. (\ref{eq:matele}), $\Xi=\xi e^{-\kappa b}$, where $\xi$ depends on the wavefunction amplitude in the barrier, $\kappa=\sqrt{2 m \Delta_b}/\hbar$, is the decay constant of the wavefunction in the barrier, of height $\Delta_b=1.5$ eV, and $m=0.5m_e$ is the effective electronic mass in the barrier \cite{Britnell2012}.

\begin{figure} 
  \centering
  \label{figuretwo}
\includegraphics[width=0.8\linewidth]{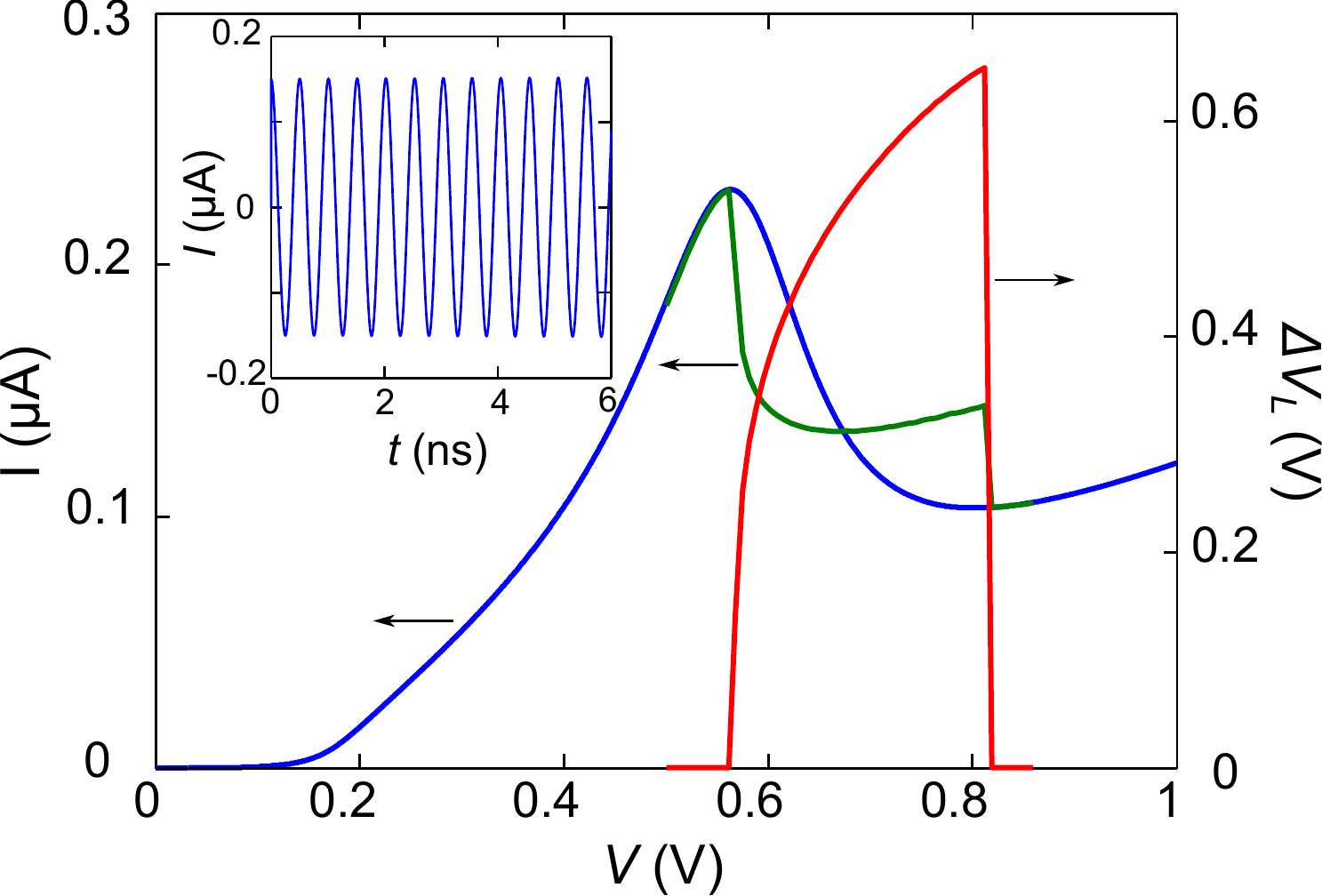} 
  \caption{Equilibrium and non-equilibrium current-voltage curves calculated for $\theta=0.9^\circ$, $L=140$ nH and $R=50$ $\Omega$. Blue curve: equilibrium current-voltage characteristic $I_b(V_b)$. Note, in equilibrium, $V_b \approx V$ and $I_b=I$. Green curve: time-averaged current $\langle I(t) \rangle_t$ vs $V$. Red curve: peak-to-peak voltage amplitude (right scale) of the stable current oscillations. Inset: $I(t)$ plot showing stable oscillations with $f=4.2$ GHz.} 
\label{fig:2} 
\end{figure}

Fig. \ref{fig:2} shows the equilibrium (static) current-voltage $I_b(V_b)$ curve (blue), where $V_b \approx V$ and $I_b=I$, calculated for an undoped device with $\theta=0.9 ^\circ$ and a barrier width $d=1.3$ nm (4 layers of hBN). These parameters are similar to those of a device which has been recently fabricated and measured \cite{Nat_Nano}. The calculated $I_b(V_b)$ curve reproduces the line-shape, position of the resonant peak and current amplitude of the measured device characteristics \cite{Nat_Comm,Nat_Nano}. The peak in current occurs when many electrons can tunnel with conservation of momentum, i.e. $\mathbf{q}-\mathbf{\Delta K}\sim 0$, corresponding to a resonant increase in the matrix element $M$, i.e. when $\phi_b=\hbar v_F \Delta K$, for $\theta$ close to $1^\circ$. Our simulations and the measurements \cite{Nat_Nano, Nat_Comm} indicate that temperature has negligible effect on the current-voltage curve when 
$V_b>kT/e \sim 30$ mV. Phonon-assisted tunneling has a relatively weak effect on the device characteristics when the graphene crystal lattices are almost aligned \cite{Nat_Nano}.

We now consider the non-equilibrium charge dynamics when the device is in a series circuit with inductance, $L$, and resistance, $R$, see Fig. \ref{fig:1}(b); the diode has its own in-built capacitance, $C$.  The primary contribution to $R$ arises from the graphene electrodes \cite{Britnell2012} and depends on the charge densities, $\rho_{B,T}$. This dependence does not have a significant effect on the high-frequency (HF) response: for most of the oscillation period, changes in $\rho_{B,T}$ do not greatly affect $R$. Therefore, to reasonable accuracy and for simplicity, we take $R$ to be independent of time. However, the value of $R$ can be changed by altering the device geometry, for example, by reducing the length of the electrodes, and we consider this effect on the performance of the GRTD. We also consider how $L$ affects the oscillation frequency, which could be controlled by careful design of the microwave circuit, for example by using a resonant cavity or integrated patch antennas \cite{Koyama2013}.

We determine the current, $I(t)$, in the contacts and external circuit by solving\cite{Greenaway2009} self-consistently the current-continuity equations:
$d\rho_{B,T}/dt=\pm (I_b-I)/A$,
where the + ($-$) sign is for the bottom (top) graphene layers, see Fig. 1(b), $\rho_{B,T}$ are related by Poisson's equation: $\epsilon F_b=\rho_B-\rho_{BD}=-(\rho_T-\rho_{TD})$,  in which $\epsilon=\epsilon_0\epsilon_r$ and $\epsilon_r=3.9$ \cite{Britnell2012,Artem_Perm} is the permittivity of the barrier, and $\rho_{BD}$ ($\rho_{TD}$) are the doping densities in each layer. The voltages across the inductor and resistor, $V_L$ and $V_R$, and the currents through them, $I_L$ and $I_R$, are given by $dI/dt=V_L/L$, $V_R=IR \label{eq:RR}$, and $V=V_R+V_{b}+V_L$.   

Following initial transient behavior, $I(t)$ either decays to a constant value or oscillates with a frequency, $f$, and time-averaged current, $\langle I(t)\rangle_t$. Fig. \ref{fig:2}, inset, shows a typical $I(t)$ curve, for $V=0.48$ V, exhibiting current oscillations with $f=4.2$ GHz. In Fig. \ref{fig:2}, we show $\langle I(t)\rangle_t$ versus $V$ (green curve) and $I_b(V_b)$ characteristics (blue curve) for an undoped device, with $\theta=0.9^\circ$, placed in a resonant circuit with $R=50$ $\Omega$ and $L=140$ nH. The plot reveals that when $V$ is tuned in the NDC region (0.55 V $<V<$ 0.8 V), $\Delta V_L=V_{L}^{{\rm max}}-V_L^{\rm min}$ (red curve) becomes non-zero indicating that self-sustained current oscillations are induced. Here, $V_L^{\text{max/min}}$ is the maximum/minimum voltage dropped across the inductor during a current oscillation period. Also, the $\langle I(t)\rangle_t$ versus $V$ curve (green) diverges from the equilibrium current, $I_b(V_b)$ (blue curve) in the NDC region. This is due to asymmetric rectification of $I(t)$ in the strongly nonlinear NDC region of $I_b(V_b)$. When the device is biased in regions of positive differential conductance (PDC), i.e. $V<0.55$ V or $V>0.8$ V, the current oscillations are suppressed and $\langle I(t)\rangle_t$ converges to $I_b(V_b)$.

This behavior is similar to that recently measured in a GRTD, where current oscillations with $f \sim$ 2 MHz were reported\cite{Nat_Nano}. That device had high circuit capacitance due to the large-area contact pads and coupling to the doped Si substrate (gate).  This effect can be modelled by placing a capacitor in parallel with the GRTD. Including this large capacitance ($65$ pF) limits the maximum frequency, as observed\cite{Nat_Nano}. Here, we consider the case when parasitic circuit capacitances are minimised, using the geometry exemplified in Fig. 1(a), so that the only significant contribution to the total capacitance is from the graphene electrodes, as described by the charge-continuity equation. This enables us to investigate the potential of GRTDs optimised for HF applications.

\begin{figure}[t!]
  \centering 
  \includegraphics[width=1.\linewidth] {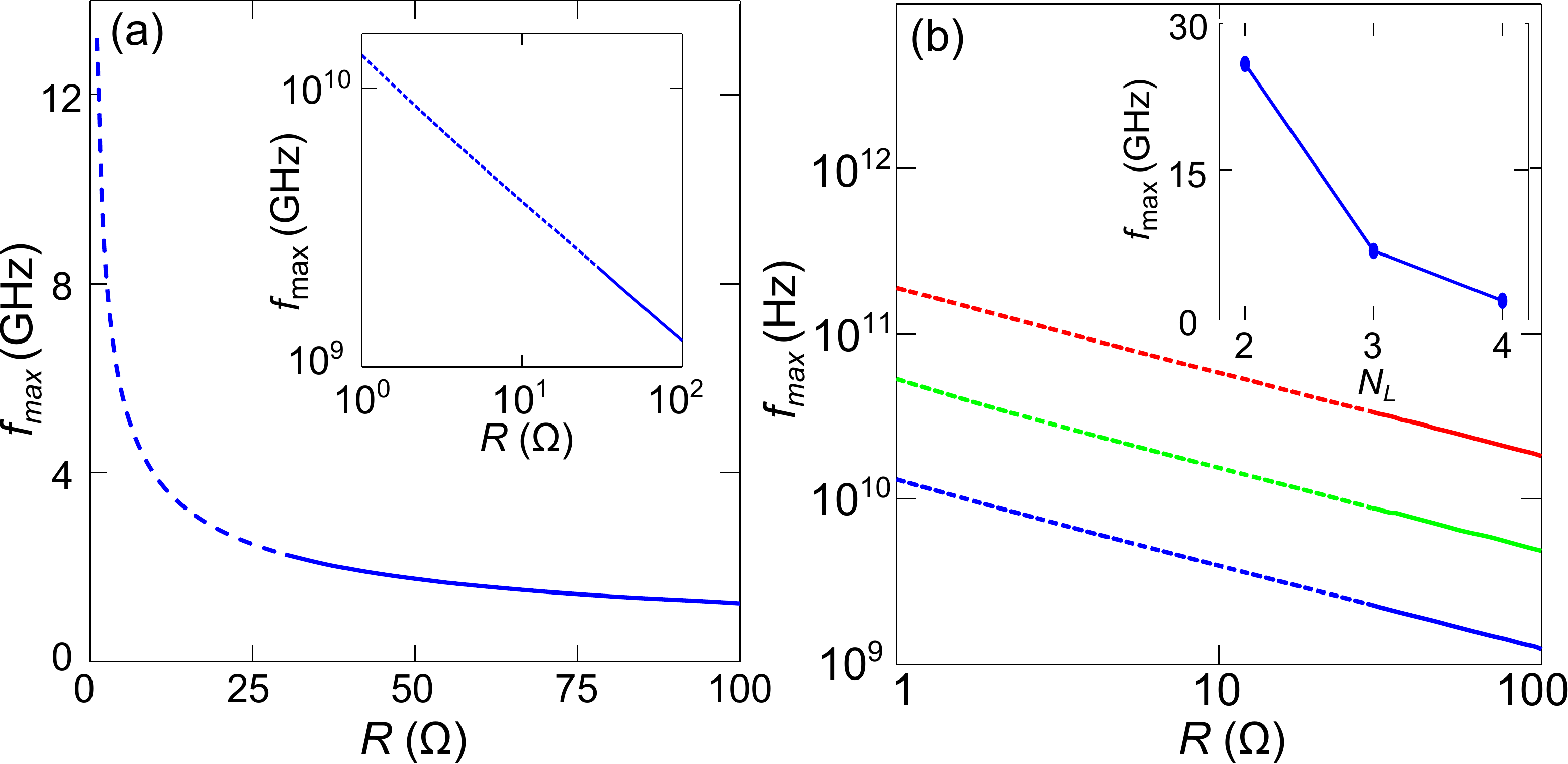}
  \caption{(a) $f_{max}(R)$ calculated when $N_L=4$. Inset: Log-log plot. (b) $f_{max}$ vs $R$ when $N_L=2$ (red), 3 (green) and 4 (blue). Inset: $f_{max}$ vs $N_L$ calculated when $R=50$ $\Omega$. Curves are shown solid over the range of $R$ presently obtainable in GRTDs and dashed for $R$ values that could be achieved by future device designs. All curves are for undoped devices.}
\label{fig:3}  
\end{figure}
A small signal analysis\cite{Hines1960} provides insight into how $L$, $R$, and the form of $I_b(V_b)$ affect the circuit response and gives an approximate oscillation frequency: 
\begin{equation}
f^s=f_0 \sqrt{\left(1-R/R_{N}\right)-Q_N^{-2}\left(1-Q_N^2 R/R_N\right)^2/4} ,
\end{equation}
where $R_N$ is the maximum negative differential resistance of the equilibrium $I(V)$ curve, the circuit factor $Q_N=R_N\sqrt{C/L}$, and $f_0=1/2\pi\sqrt{LC}$. For our device, $R_N$ is large and therefore $f^s\approx f_0$. For a given $C$ (that depends on $A$ and $d$), the oscillation frequency can be increased by reducing $L$.  The decay parameter of the small signal analysis reveals that the circuit will oscillate only if
\begin{equation}
\left( R_N/R- Q_N^2 \right)>0 .
\label{eq:decay}
\end{equation}
Consequently, $R$, and the shape of the static $I_b(V_b)$ curve are also important for optimising the HF performance. 

We now consider the fully self-consistent simulation of the charge dynamics obtained using Eqs. (1-4). Fig. \ref{fig:3}(a) shows the $f_{max}(R)$ curve calculated for the diode parameters, which compare well to recent measurements \cite{Nat_Nano}, used to produce the $I_b(V_b)$ curves in Fig. \ref{fig:2}. We determine $f_{max}(R)$ by finding the smallest $L$ value for self-sustained current oscillations. The solid part of the curve in Fig. 3(a) shows $f_{max}$ over the range of $R$ values that can be achieved by only small modifications to the design of existing devices, for example by reducing the length of the graphene between the tunnel area and the Ohmic contacts, or doping the electrodes. The dashed part of the curve is calculated for $R$ values that may be possible in future configurations. The curve reveals that for a readily-attainable $R=50$ $\Omega$, $f_{max} = 1.8$ GHz.  

Fig. 3(a), inset, reveals the power law $f_{max} \propto R^{-0.505}$, which can be derived by setting Eq. (\ref{eq:decay}) equal to zero and rearranging to find the smallest $L$ value for a given $R$, $R_N$, and $C$ \cite{Hines1960}. For this case 
\begin{equation}
f^s_{max}=(2\pi C\sqrt{R R_N})^{-1} \propto R^{-0.5},
\label{eq:fmax}
\end{equation}
which compares well with the full signal analysis.

To increase $f_{max}$, in addition to varying the external circuit parameters, we can also modify the $I_b(V_b)$ curve.  Reducing the number of hBN layers, $N_L$, in the tunnel barrier significantly increases the tunnel current  ($\sim 20 \times$ for each layer removed \cite{Britnell_Thickness}) thus reducing $R_N$ and increasing $f_{max}$, see Eq. (\ref{eq:fmax}).  Fig. \ref{fig:3}(b) shows $f_{max}(R)$ calculated for a device with $N_L=4$ (blue curve), 3 (green curve) and 2 (red curve). Reducing the barrier width produces a large gain in $f_{max}$ for all $R$.  For example, $f_{max}$ for a device with $N_L=2$ is at least an order of magnitude higher than when $N_L=4$ (e.g. for $R=50$ $\Omega$, $f_{max}$= 26 GHz when $N_L=2$, compared to $f_{max}=1.8$ GHz when $N_L=4$).

\begin{figure}
\includegraphics[width=1.\linewidth]{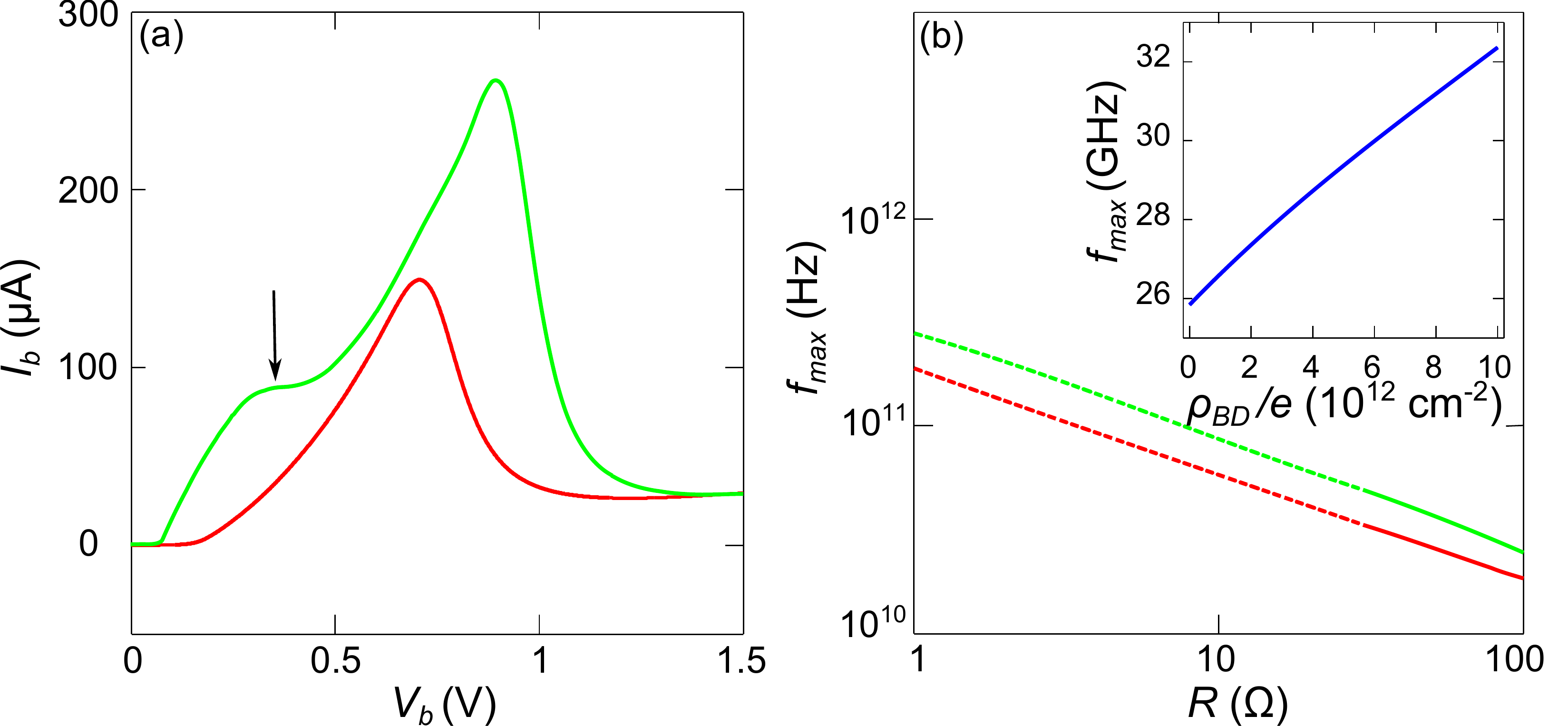} 
\caption{(a) $I_b(V_b)$ characteristics calculated for a doped (green curve, $\rho_{BD}/e=10^{13}$ cm$^{-2}$) and undoped (red curve) device, with $N_L=2$. The arrow shows the shoulder that arises due to the quantum capacitance effect. (b) $f_{max}$ vs $R$ curves for the devices in (a). Inset: $f_{max}$ vs $(\rho_{BD}/e)$ calculated when $R= 50$ $ \Omega$, with $\rho_{TD}/e=0$. Curves in (b) are shown solid over the range of $R$ presently obtainable in GRTDs and dashed for $R$ values that could be achieved by future device designs.} 
\label{fig:4}
\end{figure}

The $I_b(V_b)$ characteristics can also be modified by doping the graphene chemically \cite{doping1,doping2} or, equivalently, by applying a gate voltage, $V_g$, to shift the position of the current peak and, thereby, strongly influence $R_N$ and the peak to valley ratio \cite{Nat_Comm,Nat_Nano}. However, a gate electrode would capacitively couple to the graphene layers and the contact pads, so we do not consider it here. In Fig. \ref{fig:4}(a), we show $I_b(V_b)$ curves calculated when $N_L=2$ for an undoped (red curve) and an asymmetrically-doped device with $\rho_{BD}/e=10^{13}$ cm$^{-2}$ and $\rho_{TD}/e=0$ (green curve). When $\rho_{BD}>0$, the resonant peak occurs at higher $V_b$ than when $\rho_{BD}=0$, and the magnitude of the current peak is also higher, raising the PVR from 1.5 to 3.5. 

The shoulder of the green curve in Fig. 4(a), indicated by an arrow when $\rho_{BD}/e=10^{13}$ cm$^{-2}$, arises from the low density of states around the Dirac point. This gives rise to an additional quantum capacitance\cite{Nat_Comm,Luryi}, $C_Q$, whose effect is most prominent when the chemical potential in one layer aligns with the Dirac point in the other layer. The total device capacitance is given by 
$C^{-1}=C_G^{-1}+C_Q^{-1}$,  where $C_G=\epsilon_0\epsilon_rA/d$ is the geometric capacitance.  When $\mu_{B,T}$ passes through the Dirac point, $C_Q \rightarrow 0$ and, hence, $C \rightarrow 0$, suggesting that the RC time constant of the device could be reduced. In practice, $C_Q$ is small for only a small fraction of the oscillation period and so its effect on the fundamental frequency of $I(t)$ is negligible. 

Fig. \ref{fig:4}(b) shows the $f_{max}(R)$ curves calculated for the undoped (red) and doped (green) devices and reveals that the doped device is significantly faster for all $R$. Fig. \ref{fig:4}(b) inset shows that $f_{max}$ increases monotonically with $\rho_{BD}/e$ when $R=50$ $\Omega$; $f_{max}$ increases by a factor of 1.25 when $\rho_{BD}/e$ is increased to $10^{13}$ cm$^{-2}$ (and $f=32$ GHz) from $\rho_{BD}/e=0$ ($f=26$ GHz).  

\begin{figure} [t!]
\includegraphics[width=0.9\linewidth]{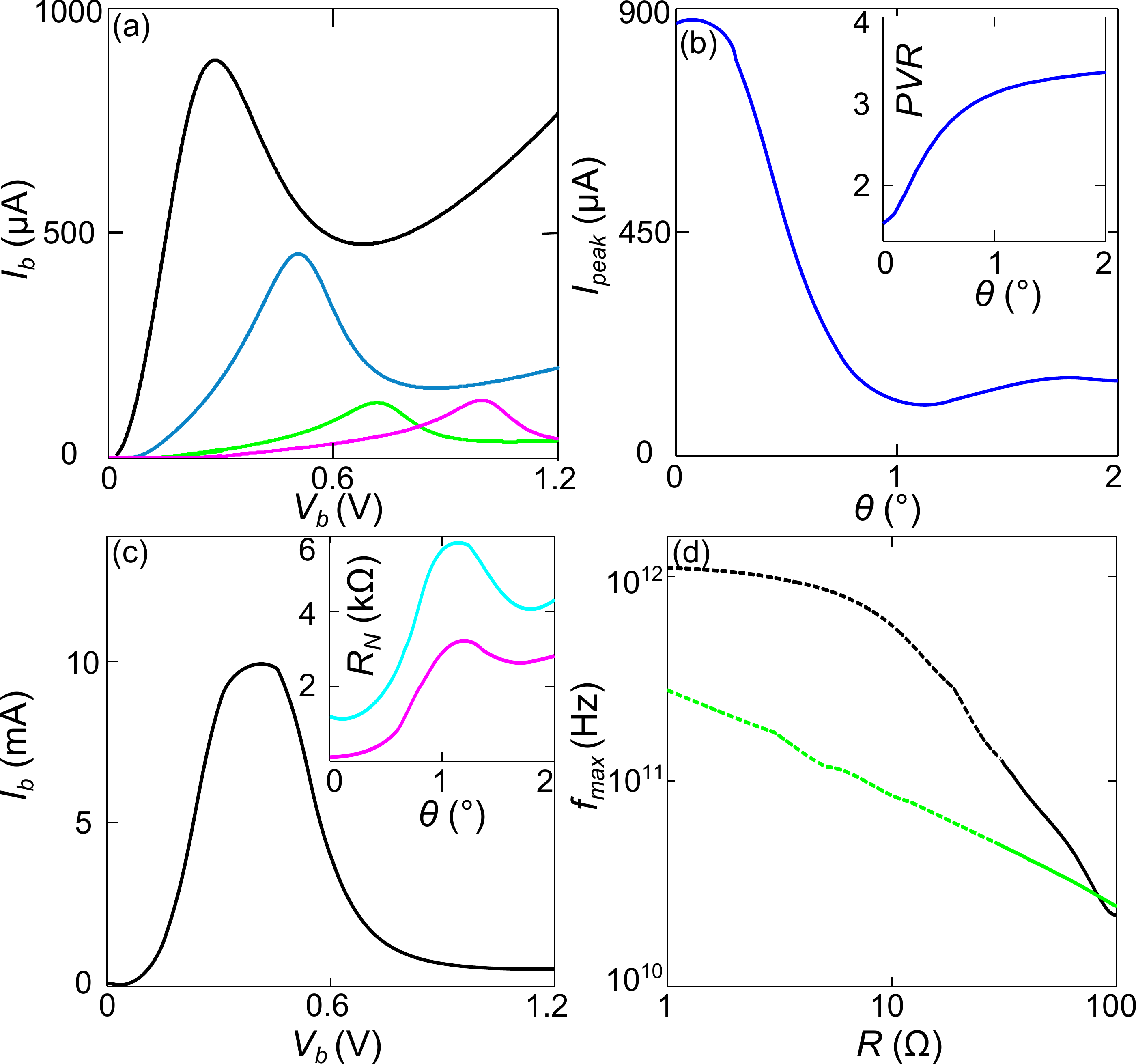}  
\caption{(a) $I_b(V_b)$ curves calculated for samples with misalignment angles $\theta=0^\circ$ (black curve), $0.5^\circ$ (blue curve), $0.9^\circ$ (green curve), $2^\circ$ (magenta curve), taking $\rho_{BD}/e=0$ cm$^{-2}$ and $N_L=2$. (b) Current amplitude at the peak vs misalignment angle, $\theta$. Inset: PVR of $I_b(V_b)$ vs $\theta$. (c) $I_b(V_b)$ calculated when $\theta=0^\circ$ and $\rho_{BD}/e=10^{13}$ cm$^{-2}$. Inset: $R_N$($\theta$) for undoped (upper curve) and $\rho_{BD}/e=10^{13}$ cm$^{-2}$ (lower curve) diodes. (d) $f_{max}$ vs $R$ curves calculated for an aligned sample (black curve) and misaligned sample with $\theta=0.9^\circ$ (green curve), when $\rho_{BD}/e=10^{13}$ cm$^{-2}$. Curves are shown solid over the range of $R$ presently obtainable in GRTDs and dashed for $R$ values that could be achieved by future device designs. For all curves, $\rho_{TD}/e=0$ cm$^{-2}$.}
\label{fig:5}
\end{figure} 

To quantify the possible benefits of lattice alignment, Fig. \ref{fig:5}(a) shows the effect of changing $\theta$ on $I_b(V_b)$. As $\theta$ increases, the position of the current peak shifts to higher $V_b$. The peak current amplitude, $I_{peak}$, decreases as $\theta$ increases due to increasing misorientation of the spatial parts of the wavefunction, see Fig. 5(b). For example, our analysis suggests that the amplitude of the resonant peak could be $\sim 10 \times$ larger for an aligned device. However, for undoped samples, the PVR increases with increasing $\theta$, see inset in Fig. 5(b), converging to a value of 3.4 as $\theta$ approaches $2^\circ$: at higher $\theta$, more states are available to tunnel resonantly at the current peak \cite{Feenstra}.  For the doped samples ($\rho_{BD}/e=10^{13}$ cm$^{-2}$), however, the valley current is close to 0 for all $\theta$, thus the PVR is consistently large, see Fig. 5(c). Consequently, the increase in current magnitude, which results from alignment, leads to higher frequencies without the reduction of power that is associated with undoped samples. We find that, generally, $R_N$ ($\propto$ $1/f_{max}^s$, see Eq. (7)) decreases with decreasing $\theta$, Fig. 5(c) inset, and with increasing $\rho_{BD}$, meaning that the oscillation frequency is highest for $\theta=0^\circ$ and when $\rho_{BD}=0$.  Fig. 5(d) shows that perfect alignment could increase $f_{max}$ by a factor of $\sim 2$, i.e. for $R=50$ $\Omega$, $f_{max}=65$ GHz when $\theta=0^\circ$ compared to $32$ GHz when $\theta=0.9^\circ$. The numerical results diverge from the small signal analysis power law of $f_{max} \propto R^{-0.5}$ as $R_N$ becomes small, see black curve of Fig. 5(d), and it becomes necessary to vary $V$ to induce current oscillations.

In conclusion, we have investigated the performance of GRTDs as the active element in RLC-based oscillators. Our simulations predict that these devices could oscillate at mid-GHz frequencies, by careful design of the RLC circuit. We have also quantified the effect of changing the parameters of the GRTD. Reducing the barrier width (a modest change to the structure of existing devices) increases the tunnel current, and thus raises the oscillation frequency by an order of magnitude. Adjustment of the doping of the electrodes can also be used to tune and enhance the oscillation frequency. Finally, we have considered the effect of misalignment of the graphene electrodes: in devices with aligned lattices, frequencies approaching 1 THz may be attainable. GaAsInAs/AlAs RTDs \cite{Suzuki2010} with two layer-thick barriers have similar peak currents and voltages as the GRTD reported here. We therefore expect that the GRTD will produce similar EM emission power ($\sim$10 $\upmu$W). Our results illustrate the potential of graphene tunnel structures for making the HF components in graphene electronics.  

\section*{Acknowledgements}

This work is supported by the EU Graphene Flagship Programme.  K.S.N. and M.T.G. acknowledge the support of the Royal Society and of The Leverhulme Trust, respectively.

\end{document}